\documentclass{article}

\usepackage{PRIMEarxiv}

\usepackage[utf8]{inputenc} 
\usepackage[T1]{fontenc}    
\usepackage{hyperref}       
\usepackage{url}            
\usepackage{booktabs}       
\usepackage{amsfonts}       
\usepackage{nicefrac}       
\usepackage{microtype}      
\usepackage{lipsum}
\usepackage{fancyhdr}       
\usepackage{graphicx}       
\graphicspath{{media/}}     
\usepackage{amsmath}
\usepackage[most]{tcolorbox}
\definecolor{bg}{RGB}{240,240,240}

\title{ChatGPT Informed Graph Neural Network for Stock Movement Prediction
}

\author{
  Zihan Chen \\
  Stevens Institute of Technology \\
  Hoboken, New Jersey, USA \\
  \texttt{zchen61@stevens.edu} \\
   \And
  Lei (Nico) Zheng \\
  Stevens Institute of Technology \\
  Hoboken, New Jersey, USA \\
  \texttt{lzheng9@stevens.edu} \\
   \And
  Cheng Lu \\
  Stevens Institute of Technology \\
  Hoboken, New Jersey, USA \\
  \texttt{clu12@stevens.edu} \\
   \And
  Jialu Yuan \\
  University of California, Los Angeles \\
  Los Angeles, California, USA \\
  \texttt{jialuy@ucla.edu} \\
   \And
  Di Zhu \thanks{Corresponding author.} \\
  Stevens Institute of Technology \\
  Hoboken, New Jersey, USA \\
  \texttt{dzhu1@stevens.edu} \\
}

\begin{document}
\maketitle

\begin{abstract}
ChatGPT has demonstrated remarkable capabilities across various natural language processing (NLP) tasks. However, its potential for inferring dynamic network structures from temporal textual data, specifically financial news, remains an unexplored frontier. In this research, we introduce a novel framework that leverages ChatGPT's graph inference capabilities to enhance Graph Neural Networks (GNN). Our framework adeptly extracts evolving network structures from textual data, and incorporates these networks into graph neural networks for subsequent predictive tasks. The experimental results from stock movement forecasting indicate our model has consistently outperformed the state-of-the-art Deep Learning-based benchmarks. Furthermore, the portfolios constructed based on our model's outputs demonstrate higher annualized cumulative returns, alongside reduced volatility and maximum drawdown. This superior performance highlights the potential of ChatGPT for text-based network inferences and underscores its promising implications for the financial sector.\footnotetext{The dataset can be accessed at our \href{https://github.com/ZihanChen1995/ChatGPT-GNN-StockPredict}{Github Repo} or https://github.com/ZihanChen1995/ChatGPT-GNN-StockPredict}
\end{abstract}

%

\section{Introduction}

The task of predicting stock price movements stands as one of the most intricate and elusive challenges in modern times. The potential for substantial investment gains underscores the urgent necessity of achieving accurate predictions \cite{de2018advances}. Owing to the efficient market hypothesis, stock prices are assumed to encapsulate all relevant market information \cite{fama1965behavior, fama1995random}. This makes the process of distinguishing genuine signals from noise an intricate endeavor that can severely impact forecasting efficacy. The academic community has responded to this challenge by formulating a wide array of statistical and machine learning models that exploit diverse features such as historical prices, news items, and market events for forecasting purposes \cite{kim2000genetic, ding2015deep, gunduz2017intraday, bustos2020stock}. However, these approaches often fail to fully recognize and incorporate the latent inter-dependencies among different equities, thus curtailing their potential for generating accurate predictions.


The complexity of forecasting stock price movements is further compounded when considering these latent inter-dependencies among equities. Two primary challenges are: 1) identifying the companies that have relevance, and 2), modeling how information permeates through them. The stock price of a company can be viewed as a synergy of the stock prices of related companies that share certain relationships with the focal company (e.g., competitors, substitutes, suppliers, etc.) \cite{chan2003stock, hoberg2016text}. Moreover, the propagation of external events can have varying impact speeds on different relevant companies, giving rise to a phenomenon called "lead-lag effect" \cite{o1999cross}. Despite efficient identification and modeling being critical, existing methods pose many limitations in capturing dynamic relationships and modeling market evolution (details will be discussed in the Related Work section).

Large Language Models (LLMs), such as ChatGPT, have garnered considerable scholarly attention since their introduction. While their applications in the expansive financial economics domain are still in a nascent stage, LLMs have demonstrated remarkable performance across a wide range of Natural Language Processing (NLP) tasks \cite{zhao2023survey, bang2023multitask}. One key factor contributing to ChatGPT's success is its extensive knowledge of entities (e.g., companies, people, events) and their relationships, which are acquired through training on massive datasets. Therefore, leveraging LLMs to automatically extract latent relationships between companies may be more efficient than manual extraction or extraction with handcrafted features \cite{hoberg2016text}. 

In the study, we present a novel approach that exploits large language models, specifically ChatGPT, to predict stock price movement. Our approach begins with employing ChatGPT to identify and extract latent inter-dependencies among equities, the results of which yield a dynamic, evolving graph that undergoes daily updates. Following this, a Graph Neural Network (GNN) is employed to generate embeddings for the target companies. The resultant embeddings are then integrated with a Long Short-Term Memory (LSTM) model to forecast stock movements for the upcoming trading day.

We evaluate the proposed model's performance using a real-world dataset, setting the DOW 30 companies as our targets. Given the last update to the DOW 30 composition in August 2020, we choose the period from September 1, 2020, to December 30, 2022, as our target period in order to capture contemporary market trends. To prevent potential data leakage issues, considering that the ChatGPT model was trained on data available only up to September 2021, we designate the test period to begin from October 1, 2021. In the task of stock movement forecasting, the experimental results demonstrate that our model consistently surpasses all baseline models in weighted F1, Micro, and Macro F1 metrics with a minimum improvement of 1.8\%. Moreover, we leverage the output of our model to construct portfolios using both long-only and long-short strategies. The evaluation of portfolio performance indicates that our model consistently exceeds benchmarks in terms of cumulative returns during the out-of-sample period. Our model also manifests a lower annualized volatility and a reduced maximum drawdown. Both results in stock movement forecasting and portfolio performance evaluation underscore the effectiveness of our ChatGPT-informed GNN model, highlighting the promising implications of LLMs for financial data processing.

This paper offers two salient contributions. First, to the best of our knowledge, this is the first study of ChatGPT’s capacity to infer network structures from textual data in the financial economics area. While ChatGPT’s robust proficiency across various NLP tasks has been well established in the existing literature \cite{katz2023gpt, zhao2023survey}, our work distinguishes itself by pioneering the connection between time-series textual data and dynamic network structures. The subsequent integration of the ChatGPT-informed network structures with GNNs also harnesses the power of deep learning models when processing large-scale, streaming datasets. Second, our experimentation with a real-world dataset provides compelling evidence of our model’s superior performance in stock movement forecasting. By constructing a portfolio based on our model’s outputs, the back-testing results consistently exhibit a higher annualized return, coupled with lower volatility and drawdown. The complexity of the stock market arises from the intricate interplay of numerous interconnected factors, such as economic indicators, the financial standing of corporations, and investor sentiment. Such intertwined dynamics render stock movement prediction a formidable task. Given that previous research has showcased how marginal advancements in predictive accuracy can translate into significant profit increments \cite{de2018advances,cheng2022financial}, the heightened performance of our model underscores its substantial practical implications in the broader financial arena.

The remainder of this paper is organized as follows. The next section provides an overview of related work on stock movement prediction, large language models, and graph neural networks. We then delve into the details of our proposed model, discussing the network structure inference using ChatGPT and the process of incorporating ChatGPT’s network outputs with GNN. Subsequently, we present our experimental setup and results. We conclude the paper by highlighting potential limitations and suggesting directions for future research.

\section{Related Work}
The forecasting of financial time series, especially in relation to stock movement prediction, has emerged as a major challenge nowadays. The ability to accurately predict stock movement is of paramount importance in shaping investment decisions, controlling financial risks, devising effective trading strategies, and comprehending the intricacies of the overall market. Despite its criticality, this forecasting task presents substantial difficulties for both researchers and practitioners. As per the Efficient Market Hypothesis \cite{fama1965behavior}, stock prices reflect all accessible information pertaining to the equity, encapsulating its historical prices, corporate events, and relevant news. Conversely, the theory of random walks postulates that future prices are as unpredictable as a series of accumulated random fluctuations \cite{fama1995random}. Consequently, the abundance of intricate data coupled with the inherent unpredictability introduces a substantial difficulty in distinguishing meaningful signals from random noises for effective predictions.

Over the years, scholars have utilized a wide variety of methods and data sources to model stock movement. Traditional statistical approaches, including linear regression, auto-regression (AR), moving average (MA), ARIMA, and GARCH, have been extensively employed for financial time series forecasting \cite{bustos2020stock}. Beyond these conventional statistical methods, machine learning techniques such as k-nearest neighbors (KNN), support vector machine (SVM), random forest, and deep learning-based methods are gaining significant traction owing to their superior predictive capacities \cite{kim2000genetic, ding2015deep, gunduz2017intraday}. In addition to modeling the relationship between historical and future prices, researchers have integrated alternate data sources like news articles, social media data, and financial reports for enhanced prediction \cite{hagenau2013automated}. However, these techniques fall short of capturing the latent inter-dependencies of stocks, thereby limiting their predictive potential.

Accurately predicting stock price movement becomes more intricate when considering the latent inter-dependencies of equities. The fluctuation of one stock can significantly impact the movement of other related stocks \cite{chan2003stock}. These relationships between stocks may manifest themselves in various ways. For instance, companies could be competitors or substitutes. For example, the bankruptcy of Silicon Valley Bank instigated a downward spiral in many bank stocks due to investor apprehension about systemic risks in the financial sector \cite{yousaf2023responses}. Alternatively, these connections between equities could stem from companies sharing supply chains. For example, the rise of ChatGPT and Microsoft's investment in OpenAI led to a surge not only in Microsoft's stock price but also in associated upstream and downstream companies like NVIDIA and Intel \cite{2023NvidiaSharesSurge}. Furthermore, given the varying degrees of inter-dependency between companies, an event may influence a set of stocks at different speeds, a phenomenon known as the lead-lag effect \cite{o1999cross}. For example, an event like "Developers file a lawsuit against Microsoft over intellectual property" would immediately impact Microsoft's stock price and gradually affect other IT companies utilizing user-generated data to train for-profit machine learning algorithms \cite{gralla2023ThisLawsuitMicrosoft}.

In an effort to capture the intricate interconnections among equities, researchers have proposed the use of Graph Neural Networks (GNN) to consolidate market information across stocks. GNN represents a novel branch of deep learning methods grounded in graph theory, wherein companies serve as nodes, and links are established between two companies sharing certain relationships \cite{zhang2019graph}. By propagating information across the network, GNN enables each node in the graph to be aware of its context, encompassing neighboring nodes and their properties \cite{wu2020comprehensive}. This leads to more effective learning and representation of the market data. For instance, Cheng et al. \cite{cheng2022financial} developed a multi-modality GNN for predicting stock price movements, demonstrating superior performance compared to other non-graph-based deep learning methods.

\begin{figure}[t]
  \centering
  \includegraphics[width=0.86\linewidth]{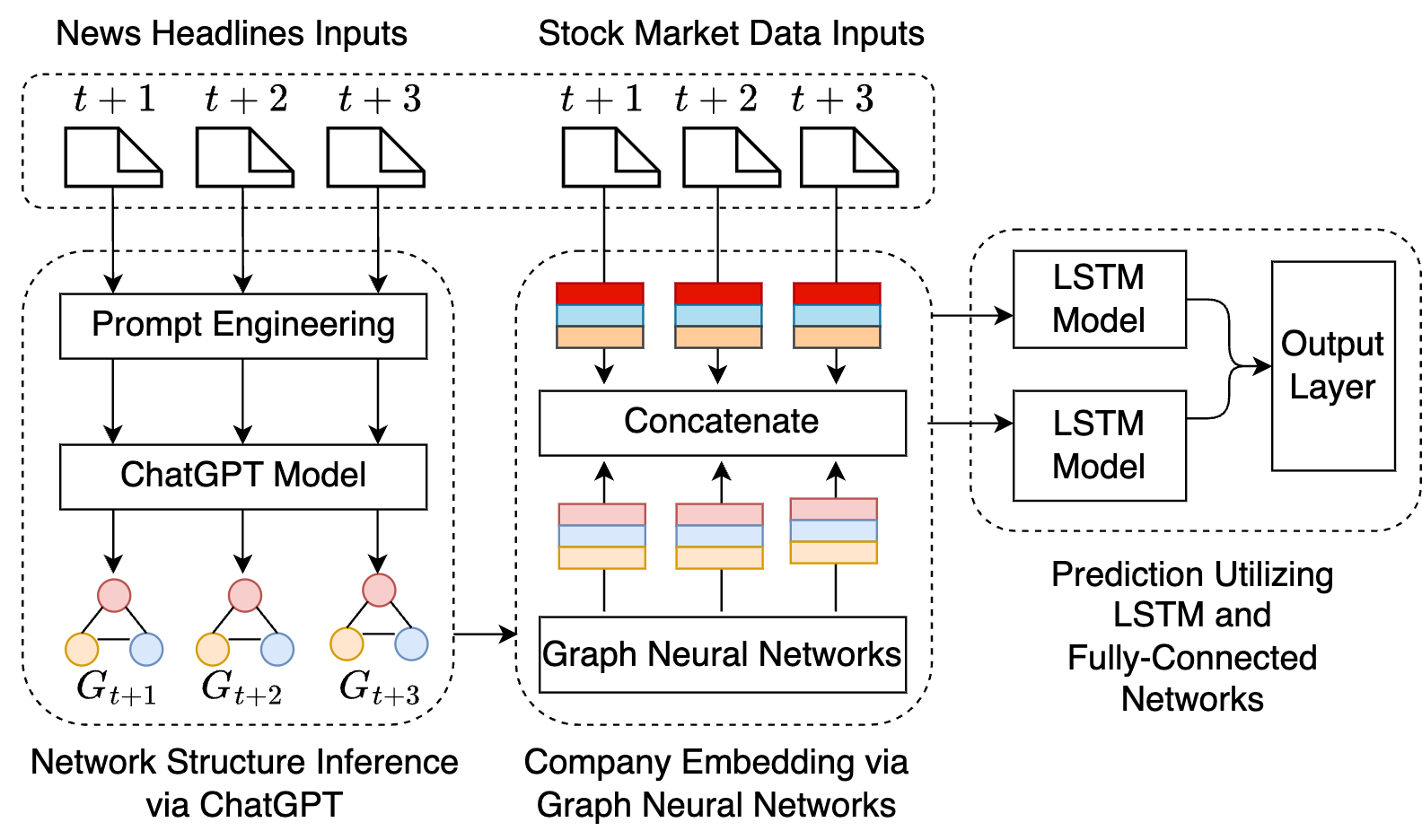}
  \caption{Framework Overview: Combining Graph Neural Network and ChatGPT to predict stock movements.}
\end{figure}

Given that GNN relies on well-defined graph structures for information propagation, accurately capturing the latent inter-dependencies among equities is crucial. Currently, two approaches are predominantly in use. The first approach involves extracting structural event tuples or leveraging text similarity from companies' business descriptions \cite{hoberg2016text} to identify company resemblances. The rationale is that companies offering similar products or services or frequently mentioned together in social media news are likely to share related stock behaviors. However, this approach may fall short of encompassing relevant domain knowledge. For instance, while both events, "David Peter leaves Starbucks" and "Steve Jobs quits Apple," pertain to employee departures, the latter would have a more profound impact on the stock market, given Steve Jobs' pivotal role in Apple. To counter this limitation, recent studies propose the integration of GNN with Financial Knowledge Graphs (FinKG) \cite{deng2019knowledge, fischer2018deep}, in which financial domain knowledge is predefined \cite{han-etal-2019-opennre, chen2020knowledge}. 
You et al. have established the efficacy and scalability of GNNs when grappling with dynamic graph structures in real-world scenarios \cite{you2022roland}.
Nevertheless, the use of predefined knowledge graphs introduces new challenges. Firstly, manually created knowledge graphs or knowledge graphs built with man-made features often fail to cover all relevant information. Secondly, as the knowledge graph is predefined, it struggles to update in a timely manner and capture emergent relationships as the market evolves. 

In our study, we propose to use Large Language Models (LLMs), such as ChatGPT, to address previously noted limitations. Although LLMs are still nascent in their application to financial economics, they have already garnered considerable scholarly interest in other areas. While not initially designed for financial data processing, LLMs have demonstrated their capability to excel in a broad spectrum of Natural Language Processing (NLP) tasks, ranging from language translation to text summarization, question answering, sentiment analysis, and text generation \cite{bang2023multitask}. Recent research has illuminated the value of these models in the financial realm. For example, Yang and Menczer \cite{yang2023large} reveal the utility of ChatGPT in distinguishing credible news sources. Similarly, Lopez-Lira and Tang \cite{lopez2023can} indicate a robust correlation between the sentiment ChatGPT generated for news headlines and the ensuing daily stock market returns.

A key element in ChatGPT's success is it has learned extensive knowledge concerning entities (such as companies, individuals, and events) and their relationships from massive training datasets. Additionally, by utilizing an attention mechanism and undergoing fine-tuning via Reinforcement Learning from Human Feedback (RLHF) \cite{ouyang2022training}, ChatGPT can better comprehend the context of textual input and identify relationships among targeted entities. These distinctive characteristics render ChatGPT an ideal tool for automatically identifying latent inter-dependencies among equities and constructing stock networks/graphs.

The utilization of ChatGPT to construct these graphs offers several advantages over previous methods \cite{hoberg2016text, deng2019knowledge, fischer2018deep} for network construction :

\begin{enumerate}
    \item[1] ChatGPT can deduce relationships between target entities from any textual input, which facilitates the use of more comprehensive and up-to-date data sources such as financial news, social media data, and corporate reports. 
    \item[2] As the relationships of interest are not confined to a predefined set of keywords, ChatGPT can recognize a broader range of relationships among companies, extending beyond shared business services and supply chains.
    \item[3] While fine-tuning is not available for ChatGPT or later versions, the method we propose can be generalized to other released versions of LLMs such as InstructGPT, Large Language Model Meta AI (LLaMA), Low-rank Adaptation (LoRA), among others. This adaptability enables more accurate applications and domain-specific customization.
\end{enumerate}

\section{Method}

Our objective is to predict the stock movement (up, down, or neutral) for a set of target companies on the next trading day. Suppose we have a total of $N$ target companies, where $i$ denotes a specific company, $t$ represents a timestamp, and $L$ corresponds to the lookback length. Accordingly, our predictive task uses features from time $t$ to $t+L$ to forecast stock movement at time $t+L+1$. To achieve this, we propose a novel framework that integrates ChatGPT and Graph Neural Network (GNN) for stock movement prediction. This framework consists of three main components: network structure inference from financial news using ChatGPT, company embedding through GNN, and stock movement prediction using sequential models and fully-connected neural networks. A comprehensive overview of our proposed framework is presented in Figure 1. We further elaborate on each component in the subsequent sections.

\subsection{Network Structure Inference via ChatGPT}

Our framework necessitates two types of time-series input features: news headlines and stock market data. The stock market data encompasses daily market information for each company, including price details (e.g., open, close, high, low), daily ask and bid, volume, and ordinary dividend amount. We use $\mathbf{S}_t$ to denote market data at time $t$, where $\mathbf{s}_{i,t}$ denotes the associated data of a specific company.
On the other hand, news headlines, sourced daily from reputable media outlets, are not company-specific and could cover various public companies. We thus exploit the inferential capabilities of ChatGPT to discern: 1) Which target companies could be affected by the day's news, and 2) How will these companies be affected: positively, negatively, or neutrally? To operationalize this, we design the following prompt for daily news headline input to ChatGPT:

\begin{tcolorbox}[enhanced jigsaw,colback=bg,boxrule=0pt,arc=0pt]

Forget all your previous instructions. I want you to act as an experienced financial engineer. I will offer you financial news headlines in one day. Your task is to:

\begin{itemize}
\item[1.] Identify which target companies will be impacted by these news headlines. Please list at least five of them.
\item[2.] Only consider companies from the target list.
\item[3.] Determine the sentiments of the affected companies: positive, negative, or neutral.
\item[4.] Only provide responses in JSON format, using the key "Affected Companies".
\item[5.] Example output: \{"Affected Companies": \{Company 1: “positive”, Company 2: “negative”\}\}
\item[6.] News Headlines are separated by "\textbackslash n"
\\
\end{itemize}

News Headlines: ...
\end{tcolorbox}

The ChatGPT response provides two insightful elements: the companies being affected by the news and their corresponding sentiment. Because prior research has demonstrated a strong association between ChatGPT's sentiment on next day's stock return \cite{lopez2023can}, we primarily focus on the "Affected Companies" output to construct a ChatGPT-Informed graph structure to feed GNN at the current stage. We build the graph $G_t=(V, E_t)$ at each timestamp by representing each target company as a node and building an edge between two companies if they were considered as "being affected together" by ChatGPT. For instance, if the "Affected Companies" output at $t$ is ['BA', 'AMGN', 'MSFT'], we construct edges $E_t$ among these ticker pairs: 'BA' – 'AMGN', 'BA' – 'MSFT', and 'AMGN' – 'MSFT'.

After gleaning these inferred relationships from news using ChatGPT, we input these graphs sequentially into a Graph Neural Network (GNN) to generate company embeddings. The GNN operation method is discussed in the next section.

\subsection{Company Embedding through GNN}

At this stage, we leverage the Graph Neural Network (GNN) to transform the nodes (companies) into vector representations. As a cutting-edge model for deep learning, GNN is adept at handling complex graph structures and embedding nodes into lower-dimensional vectors that encapsulate both nodes’ attributes and network topology \cite{zhang2019graph}. In our context of predicting stock movement, the GNN integrates the features of a company and its closely interconnected companies at a given timestamp to generate embeddings. Consequently, each company's embedding through the GNN incorporates its unique features as well as the features of relevant companies which ChatGPT considered are affected together by the news headline. Taking company $i$ and associated features at time $t$ as an example, we formally describe the GNN embeddings process as follows:

\begin{equation}
\mathbf{h}_{i,t}^{\text {GNN}}=\operatorname{GNN}\left(\mathbf{h}_{i,t} ; \mathbf{m}_{i,t} ; \Theta_{GNN}\right)
\end{equation}
where $\Theta_{GNN}$ symbolizes the trainable parameters in each layer of GNN, $\mathbf{h}_{i,t}$ denotes the original feature of company $i$, and $\mathbf{m}_{i,t}$ represents the aggregated information from its neighbors at time $t$. The final GNN embedding of company $i$ is denoted as $\mathbf{h}_{i,t}^{\text {GNN}}$.

\subsection{Sequential Models and Output Layers}

Retaining the information of the company and its neighbors, the output of the GNN is subsequently concatenated with the corresponding company's stock market data. We utilize a Long Short-Term Memory (LSTM) model as the sequential model in our framework. These combined data vectors are sequentially input into the LSTM, generating aggregated embeddings specific to each company over the lookback period. Concurrently, the stock market data undergoes a separate LSTM model to generate another set of embeddings. These two sets of embeddings are concatenated again and fed through a fully connected neural network layer to generate the final prediction for the stock movement. The process can be formalized as follows:

\begin{equation}
\mathbf{h}_{i, t}^{\text {COMB}}=\operatorname{CONCAT}\left(\mathbf{h}_{i, t}^{\text {GNN}}, \mathbf{s}_{i, t}\right)
\end{equation}
\begin{equation}
\mathbf{h}_i^{\text {COMB}}=\operatorname{LSTM}\left(\left[\mathbf{h}_{i, t}^{\text {COMB}}, \cdots \mathbf{h}_{i, t+L}^{\text {COMB}}\right] ; \Theta_{LSTM_1}\right)
\end{equation}
\begin{equation}
\mathbf{h}_i^{\text {STOCK}}=\operatorname{LSTM}\left(\left[\mathbf{s}_{i, t}, \cdots \mathbf{s}_{i, t+L}\right] ; \Theta_{LSTM_2}\right)
\end{equation}
\begin{equation}
\hat{y}_i=M L P\left(\operatorname{CONCAT}\left(\mathbf{h}_i^{\text {COMB}}, \mathbf{h}_i^{\text {STOCK}}\right) ;  \Theta_{MLP}\right)
\end{equation}
where $\Theta_{MLP}$, $\Theta_{LSTM_1}$, and $\Theta_{LSTM_2}$ are the trainable parameters. Furthermore, given that we predict stock movement at $t+L+1$, this is a classification task with three categories: up, down, and neutral. Following previous literature \cite{cheng2022financial}, we generate the category for the ground truth based on the return ($R_{i}=p_{i,t} / p_{i,t-1}-1$, where $p_i$ is the stock price) and defined thresholds ($r_{\text {up }}=0.01$, $r_{\text {down }}=-0.01$) as follows:
\begin{equation}
y_i = \begin{cases}\text { up } & R_i \geq r_{\text {up }}, \\ \text { neutral } & r_{\text {down }}<R_i<r_{\text {up } }, \\ \text { down } & R_i \leq r_{\text {down }}\end{cases}
\end{equation}

Finally, we employ cross entropy to generate the loss by comparing the predicted value with the ground truth. This loss value is then backpropagated through the model, allowing for the adjustment of trainable parameters during the iterative learning process. In the following section, we apply our proposed model to a real-world dataset to assess its performance.

\section{Experiment}

We evaluate the effectiveness of our proposed framework using a real-world dataset comprising the Dow Industrial Average 30 Companies (DOW 30) as the main subjects. Since the DOW 30 composition was last updated on August 31, 2020, we opt for the period from September 1, 2020, to December 30, 2022, as our target interval to capture the contemporary market trends. The training period extends from September 1, 2020, to September 30, 2021, consistent with the final data point integrated into ChatGPT's model training. Accordingly, the test period spans from October 1, 2021, to December 30, 2022. To gather input features for both periods, we acquire daily numerical variables of each DOW 30 company from the CRSP Databases as stock market data. For the financial news headlines, we collect 2,713,233 and 3,717,666 unique headlines for the training and test periods respectively, gleaned from 5,489 unique providers. We then extract news that not only originates from reputable media outlets but also explicitly mentions at least one DOW 30 company. This filtration process yields a refined total of 115,549 news headlines, partitioned into 50,941 for training and 64,608 for testing.

In recognition of the temporal sensitivity of news and its lag effect on the stock market, we meticulously align the news timestamp with the subsequent market period. For instance, a news headline recorded before 16:00 on Day $t$ is linked with the same day's market data, and employed to predict stock movements on Day $t+1$. Conversely, if a headline is logged after 16:00, it is assigned to the succeeding day (Day $t+1$) and used to forecast stock movement on the following day (Day $t+2$). This stratagem ensures the purity of out-of-sample test results,  further precluding potential data leakage.

\begin{table}[b]
\centering
  \caption{Model Performance of Stock Movement Prediction}
  \label{tab:freq}
  \begin{tabular}{lccc}
    \toprule
    Model & Weighted F1 & Micro F1 & Macro F1\\
    \midrule
    ChatGPT & 0.3970 & 0.4607 & 0.3085 \\
    News-Embed & 0.4059 & 0.4318 & 0.3425 \\
    Stock-LSTM & 0.4036 & 0.4132 & 0.3455  \\
    \textbf{Our Model} & 0.4133 & 0.4423 & 0.3529 \\
  \bottomrule
\end{tabular}
\end{table}

Our benchmark selection is rooted in the two types of input features we utilize. For stock market data, we deploy Long Short-Term Memory (LSTM) method \cite{hochreiter1997long}, renowned for its effectiveness in large scale time-series data analysis, and ARIMA model that are lauded for its skill in managing univariate time-series forecasting. To leverage the financial news headlines, we employ state-of-the-art sentence transformers to embed headlines into vectors, which are subsequently used as input to a MLP model for classification. Furthermore, corroborating previous research that affirms the predictive power of ChatGPT’s sentiment outputs for stock movements \cite{lopez2023can}, we incorporate the sentiment judgment from ChatGPT on stocks as a benchmark.

We assess our proposed model on two tasks: First, we scrutinize its performance in financial forecasting, specifically targeting stock movement classification. The evaluation metrics included weighted F1, Macro F1, and Micro F1 scores. Second, we construct a portfolio based on the model outputs, and evaluate its performance in terms of accumulated return, volatility, Sharpe ratio, and maximum drawdown. Detailed results from these experiments will be elucidated in the subsequent section.

\subsection{Financial Forecasting of Stock Movement}

The experimental results of stock movement forecasting are presented in Table 1, with two primary observations being made clear. First, our proposed model persistently outshines both the stock-LSTM and News-DL models in all three metrics, recording a minimum enhancement of 1.8\%. Notably, our model distinguishes itself from stock-LSTM by employing dynamic graph structures that ChatGPT generates from daily financial news. This suggests the potency of ChatGPT's zero-shot learning capability in inferring networks from text, thus advancing the predictive performance. Also, it is important to emphasize the inherent difficulty of accurately predicting stock movements, where marginal improvements can bring about significant additional profits. Earlier research has demonstrated a 0.005 increase in the Micro F1 score can result in a profit increase of 12\%, and a 1\% enhancement can lead to a 30\% profit surge \cite{de2018advances,cheng2022financial}. Consequently, our model offers considerable practical implications within the financial field.

Second, though past studies have emphasized the strong correlation between the sentiment outputs from ChatGPT and stock movements \cite{lopez2023can}, our findings indicate that amalgamating these outputs with graph neural networks amplifies performance. Despite ChatGPT delivering commendable Micro F1 scores, this is largely due to an inherent data imbalance during the testing phase, as the 58.5\% of stock movements were neutral. ChatGPT's predictive prowess falters when forecasting stock downtrends, with a score of 10.88\%, compared to our model's 19.46\% in this category. This pattern echoes in time-series models like ARIMA, which predominantly predict all movements as neutral. The enhanced ability of our model to forecast both upward and downward movements is instrumental in aiding investors to limit losses and maintain portfolio stability. In the following section, we will construct portfolios based on the outputs of the models and evaluate their economic performance.

\begin{figure}[h]
  \centering
  \includegraphics[width=0.8\linewidth]{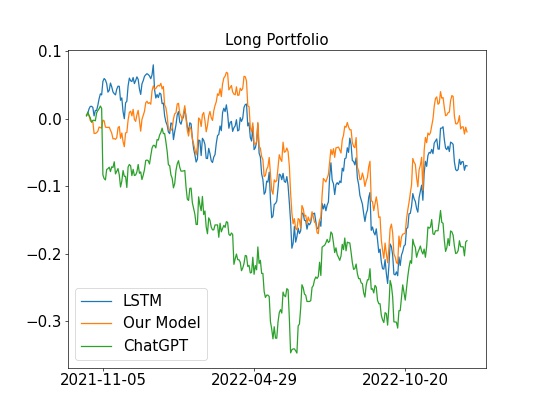}
  \caption{Comparison of Portfolio Performance During the Test Period}
\end{figure}

\subsection{Evaluation of Portfolio Performance}

We also evaluate the economic implications of our model by constructing a portfolio grounded in the model's outputs. Given that each stock's predicted outcome by our model is either upward, neutral, or downward for the next trading day, we first construct the portfolio with a long-only strategy, which lessens the exposure to risks such as short squeezes and funding liquidity. Specifically, we distribute equal investments across the stocks forecasted to rise by our model, while the remaining stocks in the portfolio are not invested in. We apply the same strategy to our proposed model and benchmarks, and conduct a backtest on these long portfolios during the out-of-sample period (October 1, 2021, to December 30, 2022). The cumulative returns for the Long Portfolio are depicted in Figure 2. As seen from the figure, our proposed model consistently outperforms both the LSTM and ChatGPT model in terms of cumulative returns. This persistent superiority signifies the effectiveness of our model in predicting positive stock returns.

Moreover, we implement a long-short strategy that forecasts both negative and positive stock returns in order to construct a self-financing portfolio. The outcomes reveal that our proposed model persistently surpasses baselines. Notably, the portfolio derived from ChatGPT outputs exhibits significantly higher annualized volatility (23.61\%) compared to our model (14.06\%). The maximum drawdown of the ChatGPT model (0.2112) also substantially exceeds that of our model (0.1242). As previously noted, this discrepancy is primarily due to ChatGPT's limitations in predicting negative returns, thereby rendering it prone to higher volatility.

In summary, our proposed model surpasses the baseline models in the task of predicting stock movements. The results provide compelling evidence that coupling GNN with ChatGPT's capabilities of inferring network structures from financial news can notably augment the predictive capacity of a model. Additionally, portfolio construction guided by our model's output consistently delivers superior performance compared to the benchmarks. This outperformance is exhibited through increased cumulative returns, along with lower annualized volatility and maximum drawdown. The robust performance across these two areas underscores the potential real-world applicability of our model in the finance industry.

\section{Discussion}

This study introduces a novel framework that capitalizes on the graph inference capabilities of ChatGPT to augment GNN forecasting performance. In our approach, ChatGPT initially distills evolving network structures from daily financial news. These inferred networks are subsequently incorporated into the GNN to produce vector embeddings, which are subsequently used in downstream prediction tasks. We assess the efficacy of our model using real-world data from the DOW 30 companies spanning from October 2021 to December 2022. The empirical findings demonstrate that our model surpasses all benchmarks in forecasting stock movements. Moreover, when portfolios are constructed based on our model’s outputs, they showcase superior cumulative returns while simultaneously exhibiting reduced volatility and drawdowns. Our research contributes to the literature by assessing the capacity of modern Language Learning Models (LLMs) to infer network structures from text. Further, it pioneers the implementation of networks inferred by ChatGPT to enhance the capabilities of GNNs. The outperformance of our model in practical scenarios emphasizes its potential implications for the financial sector, offering new perspectives and strategies in the realm of financial engineering.

Despite, to the best of our knowledge, this is the first study that integrates ChatGPT-inferred networks with GNNs, the paper is not without its limitations. First, our model leverages stock market data and time-stamped news headlines as input features. Given that stock market dynamics are influenced by a complex web of interconnected factors (including economic indicators, corporate financial health, and investor sentiment), enhancing our model with additional input features could further boost its predictive accuracy. Similarly, our study solely utilizes the network structure inferred by ChatGPT as input for GNN. Future research could consider incorporating sentiment scores as edge attributes to further improve the model's performance.

Second, our study only utilizes basic network structures in the model. However, these structures could be upgraded to more sophisticated architectures, such as replacing LSTM with transformer-based models, or employing more advanced GNN models. It is worth noting that, due to the limited scope of our sample - the DOW 30 companies - more complex GNN structures could potentially lead to oversmoothing issues \cite{chen2020measuring}. To avoid this, future research should consider expanding the dataset to include more companies, which would synergize well with deep learning's strength in handling large datasets.

Lastly, the dataset utilized in our experiment, which ends in October 2021, was the final data point input into ChatGPT. Recent advancements in ChatGPT include browsing ability and Plugins, allowing it to interact with the most recent news and information. We posit that enriching our model with the latest financial news and market information will enhance its performance, leading to more accurate forecasts and facilitating improved informed decision-making for both researchers and practitioners.

\bibliographystyle{unsrt}  
\bibliography{references}

\end{document}